\documentclass[11pt]{article}
\usepackage{amssymb}
\usepackage{amsmath}
\usepackage[english]{babel}
\usepackage{graphicx}

\def\Journal#1#2#3#4{{#1} {\bf #2}, #3 (#4)}

\def\NIMA{{\em Nucl. Instrum. Methods} A}

\def\PLB{{\em Phys. Lett.}  B}
\def\PRL{\em Phys. Rev. Lett.}
\def\PRD{{\em Phys. Rev.} D}

\def\GaC{\em Gravitation and Cosmology}
\def\GaCS{{\em Gravitation and Cosmology} Supplement}

\def\JETPL{\em JETP Lett.}
\def\PAN{\em Phys.Atom.Nucl.}
\def\CQG{\em Class. Quantum Grav.}

\def\MPLA{{\em Mod. Phys. Lett.}  A}
\def\IJTP{\em Int. J. Theor. Phys.}
\def\NJP{\em New J. of Phys.}

\def\BWP{\em Bled Workshops in Physics}
\def\JPCS{{\em J. Phys.:} Conf. Ser.}

\def\IJMPD{{\em Int. J. Mod. Phys.}  D}
\def\CPC{\em Computer Phys. Commun.}
\def\s{{\,\rm s}}

\def\({\left(}
\def\){\right)}

\def\beq{\begin{equation}}
\def\eeq{\end{equation}}
\def\bea{\begin{eqnarray}}
\def\eea{\end{eqnarray}}

\begin{document}

    \begin{center}
        \large \textbf{Accelerator probes for new stable quarks}
    \end{center}

    \begin{center}
   K. M. Belostky$^{1}$, M. Yu. Khlopov$^{1,2,3}$, K. I. Shibaev$^{1}$

    \emph{$^{1}$National Research Nuclear University "Moscow Engineering Physics Institute", 115409 Moscow, Russia \\
    $^{2}$ Centre for Cosmoparticle Physics "Cosmion" 115409 Moscow, Russia \\
$^{3}$ APC laboratory 10, rue Alice Domon et L\'eonie Duquet \\75205
Paris Cedex 13, France}

    \end{center}

\medskip

\begin{abstract}

The nonbaryonic dark matter of the Universe can consist of new stable double charged particles $O^{--}$, bound with primordial helium in heavy  neutral O-helium (OHe)"atoms" by ordinary Coulomb interaction. O-helium dark atoms can play the role of specific nuclear interacting dark matter and provide solution for the puzzles of dark matter searches. The successful development of composite dark matter scenarios appeals to experimental search for the charged constituents of dark atoms. If $O^{--}$ is a "heavy quark cluster" $\bar U \bar U \bar U$, its production at accelerators is virtually impossible and the strategy of heavy quark search is reduced to search for heavy stable hadrons, containing only single heavy quark (or antiquark). Estimates of production cross section of such particles at LHC are presented and the experimental signatures for new stable quarks are outlined.

\end{abstract}
\section{Introduction}
The cosmological dark matter can
consist of dark atoms, in which new stable charged particles are bound by ordinary Coulomb interaction (See \cite{Levels,Levels1,mpla} for review and references).
In order to avoid anomalous
isotopes overproduction, stable particles with charge -1 (and
corresponding antiparticles), as tera-particles \cite{Glashow}, should be absent \cite{Fargion:2005xz}, so that stable
negatively charged particles should have charge -2 only.

Such stable double charged particles can hardly find place in SUSY models, but there
exist several alternative elementary particle frames, in which heavy
stable -2  charged species, $O^{--}$, are predicted:
\begin{itemize}
\item[(a)] AC-leptons, predicted
in the extension of standard model, based on the approach
of almost-commutative geometry \cite{Khlopov:2006dk,5,FKS,bookAC}.
\item[(b)] Technileptons and
anti-technibaryons in the framework of walking technicolor
models (WTC) \cite{KK,Sannino:2004qp}.
\item[(c)] and, finally, stable "heavy quark clusters" $\bar U \bar U \bar U$ formed by anti-$U$ quark of 4th
 \cite{Khlopov:2006dk,Q,I,lom} or 5th  \cite{Norma} generation.
\end{itemize}

All these models also
predict corresponding +2 charge particles. If these positively charged particles remain free in the early Universe,
they can recombine with ordinary electrons in anomalous helium, which is strongly constrained in the
terrestrial matter. Therefore cosmological scenario should provide a  mechanism, which suppresses anomalous helium.
There are two possibilities, requiring two different mechanisms of such suppression:
\begin{itemize}
\item[(i)] The abundance of anomalous helium in the Galaxy may be significant, but in the terrestrial matter
there exists a recombination mechanism suppressing this abundance below experimental upper limits \cite{Khlopov:2006dk,FKS}.
\item[(ii)] Free positively charged particles are already suppressed in the early Universe and the abundance
of anomalous helium in the Galaxy is negligible \cite{mpla,I}.
\end{itemize}
These two possibilities correspond to two different cosmological scenarios of dark atoms. The first one is
realized in the scenario with AC leptons, forming neutral AC atoms \cite{FKS}.
The second assumes charge asymmetric case with the excess of $O^{--}$, which form atom-like states with
primordial helium \cite{mpla,I}.

If new stable species belong to non-trivial representations of
electroweak SU(2) group, sphaleron transitions at high temperatures
can provide the relationship between baryon asymmetry and excess of
-2 charge stable species, as it was demonstrated in the case of WTC
\cite{KK,KK2,unesco,iwara}.

 After it is formed
in the Standard Big Bang Nucleosynthesis (SBBN), $^4He$ screens the
$O^{--}$ charged particles in composite $(^4He^{++}O^{--})$ {\it
O-helium} ``atoms'' \cite{I}.

In all the proposed forms of O-helium, $O^{--}$ behaves either as lepton or
as specific "heavy quark cluster" with strongly suppressed hadronic
interaction. Therefore O-helium interaction with matter is
determined by nuclear interaction of $He$. These neutral primordial
nuclear interacting objects contribute to the modern dark matter
density and play the role of a nontrivial form of strongly
interacting dark matter \cite{McGuire:2001qj,Starkman}.

The cosmological scenario of O-helium Universe allows to explain many results of experimental searches for dark matter \cite{mpla}. Such scenario is insensitive to the properties of $O^{--}$, since the main features of OHe dark atoms are determined by their nuclear interacting helium shell. It challenges direct experimental search for the stable charged particles at accelerators and such search strongly depends on the nature of $O^{--}$.

Stable $-2$ charge states ($O^{--}$) can be elementary like AC-leptons or technileptons,
or look like elementary as technibaryons. The latter, composed of techniquarks, reveal their structure at much higher energy scale and should be produced at LHC as
elementary species. They can also be composite like "heavy quark
clusters" $\bar U \bar U \bar U$ formed by anti-$U$ quark in one of the models of fourth
generation \cite{Q,I} or $\bar u_5 \bar u_5 \bar
u_5$ of (anti)quarks $\bar u_5$ of stable 5th family in the approach
\cite{Norma}.

In the context of composite dark matter scenario accelerator search for stable particles
acquires the meaning of critical test for existence of
charged constituents of cosmological dark matter.

The signature  for AC leptons and techniparticles is unique and distinctive what  allows
to separate them  from other hypothetical exotic particles.
In particular, the ATLAS
detector has an unique potential to identify these particles and measure their masses.

Test for composite $O^{--}$ can be only indirect:
through the search for heavy hadrons, composed of single $U$ or
$\bar U$ and light quarks (similar to R-hadrons).
 Here we
study a possibility for experimental probe of this hypothesis.

\section{\label{quarks} New stable generations}
%\subsection{\label{4generation} Stable particles of 4th generation matter}
Modern precision data
on the parameters of the Standard model do not exclude \cite{Maltoni:1999ta} the existence of
the  4th generation of quarks and leptons.

In one of the approaches the 4th generation follows from heterotic string phenomenology and
its difference from the three known light generations can be
explained by a new conserved charge, possessed only by
its quarks and leptons \cite{Q,I,lom,Belotsky:2000ra}. Strict conservation of this charge makes the
lightest particle of 4th family (neutrino) absolutely
stable, but it was shown in \cite{Belotsky:2000ra} that this neutrino cannot be the dominant form of the dark matter.
The same conservation law requires the lightest quark to be long living
\cite{Q,I}. In principle the lifetime of $U$ can exceed the age of the
Universe, if $m_U<m_D$ \cite{Q,I}.

In the current implementation of the "{\it spin-charge-family-theory}" \cite{Norma} there are predicted two sets with four generations each, so that the 4th generation is unstable, while the lightest (5th generation) of the heavy set has no mixing with light families and thus is stable. If $m_{u_5}<m_{d_5}$ and their mass difference is significant, OHe dark matter cosmological scenario can be realized in this theory. For the lower possible mass scale ($\sim 1 TeV$) for the 5th generation particles, their search at LHC is possible along the same line as for stable particles of 4th generation in the approach \cite{Q,I,lom,Belotsky:2000ra}. In the successive discussion we'll consider stable $u$-type quark without discrimination of the cases of 4th and 5th generation, denoting the stable quark by $U$.

 Due to their Coulomb-like QCD attraction ($\propto \alpha_{c}^2 \cdot m_U$, where $\alpha_{c}$ is the QCD constant) stable double and triple $U$ bound states $(UUq)$, $(UUU)$  can exist
\cite{Q,Glashow,Fargion:2005xz,I,lom,Norma}. The corresponding antiparticles can be formed by heavy antiquark $\bar U$. Formation of these double and
triple states at accelerators and in
cosmic rays is strongly suppressed by phase space constraints, but they can be formed in early
Universe and strongly influence cosmological evolution of 4th
generation hadrons. As shown in \cite{I}, \underline{an}ti-
\underline{U}-\underline{t}riple state called \underline{anut}ium
or $\Delta^{--}_{3 \bar U}$ is of a special interest. This stable
anti-$\Delta$-isobar, composed of $\bar U$ antiquarks can be bound with $^4He$ in atom-like state
of O-helium \cite{Khlopov:2006dk}.

Since simultaneous production of three $U \bar U$ pairs and
their conversion in two doubly charged quark clusters $UUU$
is suppressed, the only possibility to test the
models of composite dark matter from 4th (or 5th) generation in the collider experiments is a search for production of stable hadrons containing single $U$ or $\bar U$.
$U$-quark can form lightest $(Uud)$ baryon and $(U \bar u)$ meson
 with light
quarks and antiquarks. $\bar U$ can form the corresponding stable antiparticles, like $\bar U \bar u \bar d$ and $\bar U u$. Search for these stable hadrons is similar to the R-hadrons search. The main task will be to distinguish R-hadrons from hadrons, containing quarks of 4th or 5th generation. R-hadrons will be accompanied by supersymmetric particles, what is not the case for 4th or 5th generation hadrons.

\section{\label{accelerators} Signatures for $U$-hadrons in accelerator experiments}

In spite of that the mass of $U$-quarks can be quite close to that of $t$-quark, strategy of their search should be completely different. $U$-quark in framework of the considered models is stable and will form stable hadrons at accelerator contrary to $t$-quark.

\begin{figure}
\begin{center}
\includegraphics[scale=0.5]{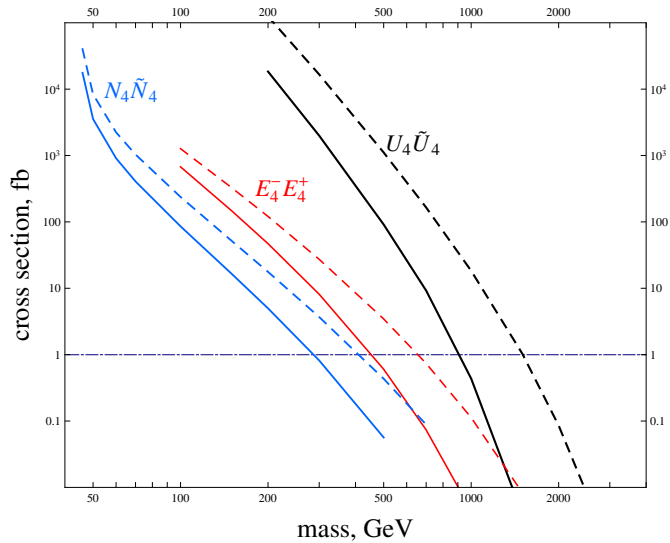}
\end{center}
\caption{Cross sections of production of 4th generation particles (N, E, U (D)) at LHC. Solid and dashed curves correspond to c.m. energies 7 and 14 TeV respectively.
Horizontal dashed line shows approximate level of sensitivity to be reached in 2012 (at the energy 7 TeV).}
\label{crossec}
\end{figure}

Detailed analysis of possibility of $U$-quark search requires quite deep understanding
of physics of interaction between (meta-)stable U-hadrons and nucleons of matter.
However, methodic for $U$-quark search can be described in general, if we
know mass spectrum of $U$-hadrons and (differential) cross sections of their production.
Cross section of $U$-quark production in pp-collisions
is presented on the Fig. \ref{crossec}. For comparison, cross sections of 4th generation
leptons are shown too. Cross sections have been calculated with program CompHEP \cite{CompHEP}. Cross sections of $U$- and $D$- quarks virtually do not differ.
For quarks ($U$ and $D$) the obtained values were re-scaled in correspondence of estimations done with program Hathor \cite{Hathor}.
Heavy stable quarks will be produced with high transverse
momentum $p_T$ and velocity, which is less than speed of light. In general,
simultaneous measurement of velocity and momentum provides us
information about mass of particle. Information on ionization
losses are, as a rule, not so good. All these features are
typical for any heavy particle, while there can be subtle
differences in the shapes of their angle- and $p_T$-distribution,
defined by concrete model, which it predicts. It is the peculiarity of
long-lived hadronic nature what can be of special importance for
clean selection of events of $U$-quarks production.

$U$-quark can form a
whole class of $U$-hadron states which can be considered as stable
in the conditions of an accelerator experiment contrary to their relics in Universe.
But in any case, as we pointed out, double and triple $U$-hadronic states cannot
be virtually created at collider. Many other hadronic states, whose
lifetime exceeds $\sim 10^{-7} \s$, should also look as stable in accelerator experiment. In the
Table 1 expected mass spectrum of $U$-hadrons, obtained with the help of
code PYTHIA \cite{pythia}, is presented.

\begin{table}
\caption{Mass spectrum and relative yields in LHC for U-hadrons}
\begin{tabular}{|p{1.3in}|p{1.1in}|p{1.2in}|p{0.55in}|} \hline
  & {\small Difference of masses of U-hadron and U-quark, GeV} & \multicolumn{2}{|p{1.9in}|}{\small Expected yields in \% (in the right columns the yields of long-lived stated are given)} \\ \hline
$\left\{U\tilde{u}\right\}^{0} ,\left\{U\tilde{d}\right\}^{+} $ & 0,330  & \multicolumn{2}{|p{1.8in}|}{39,5(3)\%, 39,7(3)\%} \\ \hline
$\left\{U\tilde{s}\right\}^{+} $ & 0,500  & \multicolumn{2}{|p{1.8in}|}{11,6(2)\%} \\ \hline
$\left\{Uud\right\}^{+} $ & 0,579  & 5,3(1)\% &  7,7(1)\% \\ \hline
$\left\{Uuu\right\}_{1}^{++} ,\left\{Uud\right\}_{1}^{+} ,$ $\left\{Udd\right\}_{1}^{0} $ & 0,771 & 0,76(4)\%, 0,86(5)\%, 0,79(4)\% &  \\ \hline
$\left\{Usu\right\}^{+} ,\left\{Usd\right\}^{0} $ & 0,805 & 0,65(4)\%, 0,65(4)\% &  1,51(6)\% \\ \hline
$\left\{Usu\right\}_{1}^{+} ,\left\{Usd\right\}_{1}^{0} $ & 0,930 & 0,09(2)\%, 0,12(2)\% &  \\ \hline
$\left\{Uss\right\}_{1}^{0} $ & 1,098 & \multicolumn{2}{|p{1.8in}|}{0,005(4)\%} \\ \hline
\end{tabular}
\end{table}

The lower indexes in notation of $U$-hadrons in the Table 1 denote the nonzero spin $s=1$ of the pair of light quarks.
From comparison of masses of different $U$-hadrons
it follows that all $s=1$ $U$-hadrons decay quickly emitting $\pi$-meson or $\gamma$-quantum,
except for $(Uss)$-state. In the right column the expected relative yields are presented.
Unstable $s=1$ $U$-hadrons decay onto
respective $s=0$ states, increasing their yields.

There are two
mesonic states being quasi-degenerated in mass: $U\bar u$ and $U\bar d$
(we skip here discussion of strange $U$-hadrons).
Interaction with the medium composed of $u$ and
$d$ quarks transforms $U$-hadrons into those ones containing
$u$ and $d$ (as it is the case in the early Universe \cite{Q,I,lom}).
The created pair of $U\,\bar U$ quarks will fly
out of the vertex of pp-collision as $U$-hadrons with
positive charge in 60\% of all $U$-quark events
and as neutrals in 40\% (correspondingly, 60\% with negative charge and 40\% neutrals
for $\bar U$ hadrons).
After traveling through the matter of detectors, at a distance of a few nuclear lengths from
vertex, $U$-hadrons will transform in (roughly) 100\% of positively
charged hadrons $(Uud)$, while $\bar U$ -hadrons will convert in 50\% into
negatively charged $\bar U$ -hadron $(\bar U d)$ and in 50\% to neutral
$\bar U$ -hadron $(\bar U u)$.

This feature will enable to discriminate the considered case of $U$-quarks from variety of alternative models,
predicting new heavy stable particles.

 Note that if the mass
of Higgs boson exceeds $2m$, its decay channel into the pair of stable
$Q \bar Q$ will dominate over the $t \bar t$, $2W$, $2Z$ and
invisible channel to neutrino pair of 4th generation
\cite{nuHiggs}. It may be important for the strategy of heavy
Higgs searches.

\section{Conclusions}

%\medskip
The cosmological dark matter can be formed by
stable heavy double charged particles bound in neutral OHe dark atoms with primordial He nuclei by ordinary Coulomb interaction. This scenario sheds new light on the nature of dark matter and offers nontrivial solution for the puzzles of direct dark matter searches. It can be realized in the model of stable 4th generation or in the approach unifying spin and charges and challenges for experimental probe at accelerators.
 In the context of this scenario search for new heavy stable quarks acquires the meaning of direct experimental probe for charged constituents of dark atoms of dark matter.

The $O^{--}$ constituents of OHe in the model of stable 4th generation and in the "{\it spin-charge-family-theory}" are "heavy quark clusters" $\bar U \bar U \bar U$. Production of such clusters (and their antiparticles) at accelerators is virtually impossible. Therefore experimental test of the hypothesis of stable $U$ quark is reduced to the search for stable or metastable $U$ hadrons, containing only single heavy quark or antiquark. The first year of operation at the future 14 TeV energy of the LHC has good discovery potential for $U$($D$)-quarks with mass up to 1.5 TeV, while the level of sensitivity expected in the 2012 at the LHC energy 7 TeV can approach to the mass of 1 TeV. $U$-hadrons born at accelerator will
distinguish oneself by high $p_T$, low velocity, by effect of a charge flipping
during their propagation through the detectors. All these features
enable to strongly increase the efficiency of event selection from not only
background but also from alternative hypothesis.  In particular, we show that the detection of positively charged $U$-baryon in coincidence with $\bar U$-mesons (50\% neutrals and 50\% negatively charged) provides a distinct signature for the stable $U$ quark. Analysis of other channels of new particles production provides distinctions from the case of R-hadrons. In the latter case all the set of supersymmetric particles should be produced.

It should be noted that the "{\it spin-charge-family-theory}" predicts together with stable 5th generation also 4th generation of quarks and leptons, which are mixed with the three known families and thus unstable. Experimental probe for new unstable heavy particles implies definite prediction for their mass spectrum and branching ratios for their modes of decay.

%\bigskip

%{\centering{ \large \textbf{Acknowledgments}} }

\section {Acknowledgments}

%\medskip

We would like to thank Norma Mankoc-Borstnik, all the
participants of Bled Workshop and A.S.Romaniouk for stimulating discussions.

%\bigskip

%{\centering{ \large \textbf{References}} }

%\section*{References}

%\medskip


\begin{thebibliography}{99}

\bibitem{Levels}
  M.~Y.~Khlopov, A.~G.~Mayorov and E.~Y.~Soldatov,
  \Journal{\JPCS}{309}{012013}{2011}.

\bibitem{Levels1}
  M.~Y.~Khlopov, A.~G.~Mayorov and E.~Y.~Soldatov,
  \Journal{\BWP}{11}{73}{2010}.

%\cite{mpla}
\bibitem{mpla}
  M.~Y.~Khlopov, arXiv:1111.2838, to be published in {\em Mod. Phys. Lett.} A {\bf 26} (2011); arXiv:1111.2887, to be published in {\em Proc. ICATPP2011}.

\bibitem{Glashow} S.~L.~Glashow,
 % ``A sinister extension of the standard model to SU(3) x SU(2) x SU(2) x
 % U(1),''
  arXiv:hep-ph/0504287.
  %%CITATION = HEP-PH 0504287;%%

\bibitem{Fargion:2005xz}
  D. Fargion and M. Khlopov,
 % ``Tera-leptons shadows over sinister Universe,''
  arXiv:hep-ph/0507087.
  %%CITATION = HEP-PH 0507087;%%



%\bibitem{Legonkov}

%Gravitation and Cosmology {\bf 11}, 27 (2005); astro-ph/0504621.




%\bibitem{Okun} M. Maltoni {\it et al}, \Journal{\PLB}{476}{107}{2000};
%V.A. Ilyin {\it et al}, \Journal{\PLB}{503}{126}{2001}; V.A. Novikov
%{\it et al}, \Journal{\PLB}{529}{111}{2002};
%\Journal{\JETPL}{76}{119}{2002}.

%\cite{Khlopov:2006dk}
\bibitem{Khlopov:2006dk}
  M.~Y.~Khlopov,
%  ``New symmetries in microphysics, new stable forms of matter around us,''
  arXiv:astro-ph/0607048.
  %%CITATION = ASTRO-PH/0607048;%%


\bibitem{5} C.~A.~Stephan,
%  ``Almost-commutative geometries beyond the standard model,''
  arXiv:hep-th/0509213.
  %%CITATION = HEP-TH 0509213;%%.

\bibitem{FKS} D.~Fargion {\it et al},
\Journal{\CQG}{23}{7305}{2006};
%Class. Quantum Grav.  {\bf 23}, 7305 (2006);
  %``Cold dark matter by heavy double charged leptons?,''
%  arXiv:astro-ph/0511789.
  %%CITATION = ASTRO-PH 0511789;%%
M. Y. Khlopov and C. A. Stephan, arXiv:astro-ph/0603187.

\bibitem{bookAC} A. Connes {\em Noncommutative Geometry} (Academic Press, London and San
Diego, 1994).



%\cite{KK}
\bibitem{KK}
  M.~Y.~Khlopov and C.~Kouvaris,  \Journal{\PRD}{77}{065002}{2008}.
%  ``Strong Interactive Massive Particles from a Strong Coupled Theory,''
%  arXiv:0710.2189 [astro-ph].
  %%CITATION = ARXIV:0710.2189;%%

%\cite{Sannino:2004qp}
\bibitem{Sannino:2004qp}
F.~Sannino and K.~Tuominen,  \Journal{\PRD}{71}{051901}{2005};
%``Orientifold Theory Dynamics and Symmetry Breaking,''
% Phys.\ Rev.\ D {\bf 71}, 051901 (2005);
%arXiv:hep-ph/0405209.
%%CITATION = HEP-PH 0405209;%%
%\cite{Hong:2004td}
%\bibitem{Hong:2004td}
  D.~K.~Hong {\it et al}, \Journal{\PLB}{597}{89}{2004};
  % ``Composite Higgs from higher representations,''
%    Phys.\ Lett.\ B {\bf 597}, 89 (2004);
 % arXiv:hep-ph/0406200.
  %%CITATION = HEP-PH 0406200;%%
%\cite{Dietrich:2005jn}
%\bibitem{Dietrich:2005jn}
  D.~D.~Dietrich {\it et al}, \Journal{\PRD}{72}{055001}{2005};
  %``Light composite Higgs from higher representations versus electroweak
  %precision measurements: Predictions for LHC,''
  %Phys.\ Rev.\ D {\bf 72}, 055001 (2005);
  %arXiv:hep-ph/0505059.
  %%CITATION = HEP-PH 0505059;%%
%\cite{Dietrich:2005wk}
%\bibitem{Dietrich:2005wk}
  D.~D.~Dietrich {\it et al}, \Journal{\PRD}{73}{037701}{2006};
 % ``Light composite Higgs and precision electroweak measurements on the Z
 % resonance: An update,''
  %arXiv:hep-ph/0510217;
  %To appear in PRD.
  %%CITATION = HEP-PH 0510217;%%
%\cite{Gudnason:2006ug}
%\bibitem{Gudnason:2006ug}
  S.~B.~Gudnason {\it et al}, \Journal{\PRD}{73}{115003}{2006};
  %``Towards working technicolor: Effective theories and dark matter,''
 % Phys.\ Rev.\  D {\bf 73}, 115003 (2006);
 % arXiv:hep-ph/0603014.
  %%CITATION = PHRVA,D73,115003;%%
%\cite{Gudnason:2006yj}
%\bibitem{Gudnason:2006yj}
  S.~B.~Gudnason {\it et al}, \Journal{\PRD}{74}{095008}{2006}.
  %``Dark matter from new technicolor theories,''
  %Phys.\ Rev.\  D {\bf 74}, 095008 (2006);
  %arXiv:hep-ph/0608055.
  %%CITATION = PHRVA,D74,095008;%%
\bibitem{Q}  %K.M.Belotsky, D.Fargion, M.Yu.Khlopov, R.Konoplich, M.G.Ryskin and K.I.Shibaev, to be published in Astroparticle Physics; hep-ph/0411271;\\
K.M.Belotsky {\it et al}, \Journal{\GaC}{11}{3}{2005}
%Gravitation and Cosmology {\bf 11}, 3 (2005).

\bibitem{I} M.Yu. Khlopov, \Journal{\JETPL}{83}{1}{2006}.
%JETP Lett. {\bf 83}, 1 (2006)
%[Pisma Zh.\ Eksp.\ Teor.\ Fiz.  {\bf 83}, 3 (2006)];
%arXiv:astro-ph/0511796.
%\cite{lom}
\bibitem{lom}
  K.~Belotsky {\it et al},
  %``Stable matter of 4th generation: Hidden in the Universe and close to
  %detection?,''
  arXiv:astro-ph/0602261.
  %%CITATION = ASTRO-PH 0602261;%%
%\bibitem{KPS06}
  K.~Belotsky {\it et al}, \Journal{\GaC}{12}{1}{2006};
  %, Gravitation and Cosmology {\bf 12}, 1   (2006);
  %``Composite dark matter and its charged constituents,''
 % arXiv:astro-ph/0604518.
 %\cite{Belotsky:2008se}
%\bibitem{Belotsky:2008se}
  K.~Belotsky, M.Yu.Khlopov,K.I.Shibaev, Stable quarks of the 4th family? in Eds. N. L. Watson and T. M. Grant: "The Physics of Quarks: New Research." ( Horizons in World Physics, V.265), NOVA Publishers, Hauppauge NY, 2009, PP.19-47;
  %``,''
  arXiv:0806.1067 [astro-ph].
  %%CITATION = ARXIV:0806.1067;%%




%\cite{Norma}
\bibitem{Norma}
N.S. Manko\v c Bor\v stnik, \Journal{\BWP}{11}{105}{2010}; A. Bor\v stnik Bra\v ci\v
c, N.S. Manko\v c Bor\v stnik, \Journal{\PRD}{74}{073013}{2006};
N.S. Manko\v c Bor\v stnik, \Journal{\MPLA}{10}{587}{1995}; N.S.
Manko\v c Bor\v stnik, \Journal{\IJTP}{40}{315}{2001}; G. Bregar, M.
Breskvar, D. Lukman, N.S. Manko\v c Bor\v stnik,
\Journal{\NJP}{10}{093002}{2008}.

%\cite{KK2}
\bibitem{KK2}
  M.~Y.~Khlopov and C.~Kouvaris, \Journal{\PRD}{78}{065040}{2008}
  %``Composite dark matter from a model with composite Higgs boson,''
  %Phys.\ Rev.\  D {\bf 78} (2008) 065040
  %[arXiv:0806.1191 [astro-ph]].
  %%CITATION = PHRVA,D78,065040;%%

%\cite{unesco}
\bibitem{unesco}
  M.~Y.~Khlopov,
  %``The puzzles of dark matter searches,''
  {\it AIP Conf. Proc.}  {\bf 1241}, 388 (2010).
  %[arXiv:0911.5685 [astro-ph.CO]].
  %%CITATION = APCPC,1241,388;%%

  %\cite{iwara}
\bibitem{iwara}
  M.~Y.~Khlopov, A.~G.~Mayorov and E.~Y.~Soldatov, \Journal{\IJMPD}{19}{1385}{2010}.
  %``Composite Dark Matter and Puzzles of Dark Matter Searches,''
  %Int.\ J.\ Mod.\ Phys.\  D {\bf 19} (2010) 1385
  %[arXiv:1003.1144 [astro-ph.CO]].
  %%CITATION = IMPAE,D19,1385;%%


%\cite{McGuire:2001qj}
\bibitem{McGuire:2001qj}
B.\,D. Wandelt et al.,
 % ``Self-interacting dark matter,''
  arXiv:astro-ph/0006344;
  %%CITATION = ASTRO-PH 0006344;%%
P.\,C. McGuire and P.\,J. Steinhardt,
  %``Cracking open the window for strongly interacting massive particles as  the
  %halo dark matter,''
  arXiv:astro-ph/0105567;
  %%CITATION = ASTRO-PH 0105567;%%
 G. Zaharijas and G.\,R. Farrar, \Journal{\PRD}{72}{083502}{2005}
  %``A Window in the Dark Matter Exclusion Limits,''
 % Phys.~Rev. {\bf D72}, 083502 (2005);
 % arXiv:astro-ph/0406531.
  %%CITATION = ASTRO-PH 0406531;%%

\bibitem{Starkman}
  C.\,B. Dover {\it et al}, \Journal{\PRL}{42}{1117}{1979};
  %``Cosmological Constraints On New Stable Hadrons,''
 % Phys.~Rev.~Lett. {\bf 42}, 1117 (1979);
  %%CITATION = PRLTA,42,1117;%%
  S. Wolfram, \Journal{\PLB}{82}{65}{1979};
  %``Abundances Of Stable Particles Produced In The Early Universe,''
  %Phys.~Lett.  {\bf B82}, 65 (1979);
  %%CITATION = PHLTA,B82,65;%%
G.\,D. Starkman  {\it et al}, \Journal{\PRD}{41}{3594}{1990};
%Phys.~Rev. {\bf D41}, 3594 (1990);
 D.~Javorsek  {\it et al}, \Journal{\PRL}{87}{231804}{2001};
  %``New experimental limits on strongly interacting massive particles at the
  %TeV scale,''
  %Phys.~Rev.~Lett. {\bf 87}, 231804 (2001);
  %%CITATION = PRLTA,87,231804;%%
S. Mitra, \Journal{\PRD}{70}{103517}{2004};
  %``Uranus' anomalously low excess heat constrains strongly interacting  dark
  %matter,''
  %Phys.~Rev. {\bf D70}, 103517 (2004);
  %arXiv:astro-ph/0408341;
  %%CITATION = ASTRO-PH 0408341;%%
  G.~D.~Mack  {\it et al}, \Journal{\PRD}{76}{043523}{2007};
  %``Towards Closing the Window on Strongly Interacting Dark Matter:
  %Far-Reaching Constraints from Earth's Heat Flow,''
 % Phys.\ Rev.\  D {\bf 76} (2007) 043523
 % [arXiv:0705.4298 [astro-ph]].
  %%CITATION = PHRVA,D76,043523;%%

\bibitem{Maltoni:1999ta}
M. Maltoni {\it et al.}, \Journal{\PLB}{476}{107}{2000};
V.A. Ilyin {\it et al.}, \Journal{\PLB}{503}{126}{2001};
V.A. Novikov {\it et al.}, \Journal{\PLB}{529}{111}{2002}; \Journal{\JETPL}{76}{119}{2002}.

\bibitem{Belotsky:2000ra}
K.M.Belotsky, M.Yu.Khlopov and K.I.Shibaev, \Journal{\GaCS}{6}{140}{2000};
%\bibitem{Belotsky:2005uj}
K.M.Belotsky {\it et al.}, \Journal{\GaC}{11}{16}{2005};
%\bibitem{Belotsky:2004st}
K.M.Belotsky {\it et al.}, \Journal{\PAN}{71}{147}{2008}.

%\cite{Sannino:2004qp,Hong:2004td,Dietrich:2005jn,Dietrich:2005wk,Gudnason:2006ug,Gudnason:2006yj}

\bibitem{CompHEP}
E.Boos et al, [CompHEP Collaboration], \Journal{\NIMA}{534}{250}{2004}.
%CompHEP 4.4: Automatic computations from Lagrangians to events, Nucl. Instrum. Meth. A534 (2004) 250 %(arXiv:hep-ph/0403113).

A.Pukhov et al, CompHEP - a package for evaluation of Feynman diagrams and integration over multi-particle phase space. User's manual for version 3.3, INP MSU report 98-41/542 (arXiv:hep-ph/9908288)

Home page: http://comphep.sinp.msu.ru

\bibitem{Hathor}
M. Aliev et al., \Journal{\CPC}{182}{1034}{2011}.
%Comput.Phys.Commun.182:1034-1046,2011 (arXiv:1007.1327 (hep-ph)).

\bibitem{pythia} T. Sj\"ostrand et al., \Journal{\CPC}{135}{238}{2001}.
%Computer Phys. Commun. 135 (2001) 238 (hep-ph/0010017).

\bibitem{nuHiggs} K.M. Belotsky et al.,\Journal{\PRD}{68}{054027}{2003}.
%Phys.Rev. \textbf{D68} (2003), 054027; hep-ph/0210153.



\end{thebibliography}
\end{document}